\documentclass[pra,twocolumn,10pt,showpacs,amsmath,amssymb,footinbib,superscriptaddress,floatfix]{revtex4-1}

\usepackage[english]{babel}  
\usepackage{amsfonts,graphicx}
\usepackage{mathrsfs}

\def\bra#1{\mathinner{\langle{#1}|}}
\def\ket#1{\mathinner{|{#1}\rangle}}
\def\braket#1{\mathinner{\langle{#1}\rangle}}

{\catcode`\|=\active 
  \gdef\Braket#1{\begingroup
\mathcode`\|32768\let|\BraVert\left<{#1}\right>\endgroup}}
\def\BraVert{\egroup\,\mid\,\bgroup}

\usepackage{cancel}
\usepackage{color}
\definecolor{Blue}{rgb}{0,0,1}
\definecolor{Red}{rgb}{1,0,0}
\definecolor{Green}{rgb}{0,1,0}
\definecolor{Purp}{rgb}{.2,0,.2}
\definecolor{white}{rgb}{1,1,1}

\newcommand{\tr}{\mbox{tr}}

\newcommand{\A}{\mathcal{A}}
\newcommand{\B}{\mathcal{B}}

\begin{document}

\title{Positivity in the presence of initial system-environment correlation}

\author{Kavan Modi}
\email{kavan@quantumlah.org}
\affiliation{Department of Physics, University of Oxford, Clarendon Laboratory, Oxford, UK}
\affiliation{Centre for Quantum Technologies, National University of Singapore, Singapore}

\author{C\'{e}sar A. Rodr\'{i}guez-Rosario} 
\email{cesar.rodriguez@bccms.uni-bremen.de}
\affiliation{Department of Chemistry and Chemical Biology, Harvard University, Cambridge MA, USA}
\affiliation{Bremen Center for Computational Materials Science, University of Bremen, Bremen, Germany}

\author{Al\'an Aspuru-Guzik}
\affiliation{Department of Chemistry and Chemical Biology, Harvard University, Cambridge MA, USA}

\date{\today}

\begin{abstract}
The constraints imposed by the initial system-environment correlation can lead to nonpositive Dynamical maps. We find the conditions for positivity and complete positivity of such dynamical maps by using the concept of an assignment map. Any initial system-environment correlations make the assignment map nonpositive, while the positivity of the dynamical map depends on the interplay between the assignment map and the system-environment coupling. We show how this interplay can reveal or hide the nonpositivity of the assignment map. We discuss the role of this interplay in Markovian models.
\end{abstract}
\maketitle


\emph{Introduction.}---The open quantum systems formalism is the standard tool used to understand and model the decoherence and thermalization of quantum systems. In this formalism, the total state of the system ($S$) and its environment ($E$), described by the density matrix $\rho^{SE}$, evolves unitarily. However, the focus is only on the dynamics of the density matrix $\eta^{S}$ of $S$ by averaging the degrees of freedom of $E$. Open quantum systems are essential for physics~\cite{Angelbook}, quantum information~\cite{Nielsen00a}, for simulating chemistry~\cite{Yuen09, Yuen10}, and in ultrafast spectroscopy~\cite{Yuen11a}. In many of these fields, it is customary to assume that at the initial time the system is uncorrelated with the environment. This assumption simplifies the mathematical structure of the map. However, recently many researchers have realized that many systems of importance are initially correlated with the surroundings and have pursued investigations on  systems that admit initial correlations~\cite{PhysRevA.84.032112, PiiloNatPhys}. It is well known that a system initially correlated with its environment may suffer from nonpositive dynamics~\cite{Rodriguez:2007p123}. In this article we tackle the question of how the initial system-environment ($SE$) correlation and the $SE$ coupling affects the positivity of dynamics.

\begin{center}
\begin{figure}[t!]
\includegraphics[scale=0.4]{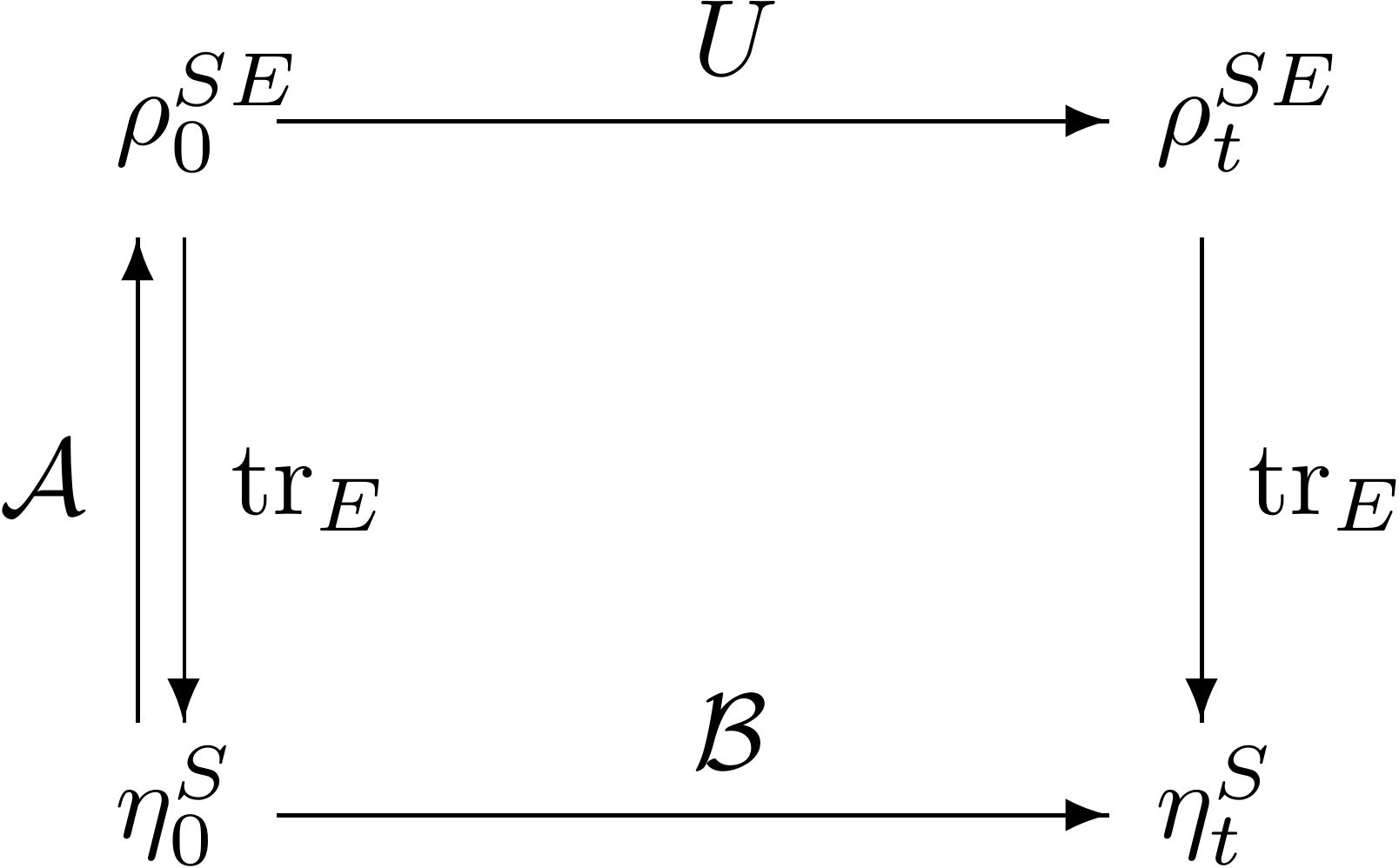}
\caption{\label{arrowdiag} \emph{Reduced dynamics from total dynamics.} The total state evolves unitarily $\rho^{SE}_t=U\rho^{SE}_0U^\dag$. The initial state of system ($S$), $\eta^S_0=\mbox{tr}_{E}(\rho_0^{SE})$, is mapped to final state $\eta^S_t=\B(\eta_0^{S})$ by the dynamical map $\B$. This process may also be seen as $\eta_0$ assigned to $\rho_0$ by the assignment map $\A$ followed by the unitary transformation $U\left( \cdot \right)U^\dagger $, and finally the environment ($E$) is traced by $\mbox{tr}_{E} $, as in Eq.~\eqref{dmap1}.}
\end{figure}
\end{center}

The \emph{dynamical map} $\mathcal{B}$ describes dynamics of the reduced system~\cite{Sudarshan:1961p2145, SudarshanJordan61, kraus}. The relationship between the total dynamics, and the dynamics of $S$ is shown in Fig.~\ref{arrowdiag}, such that the map is defined as the super-operator
\begin{gather}\label{dmap1}
\mathcal{B}\left(\eta_0 \right)=\mbox{tr}_E\left[U\rho^{SE}_0 U^\dagger\right] 
=\mbox{tr}_E\left[U\A\left[ \eta^{S}_0 \right] U^\dagger\right] = \eta^{S}_t,
\end{gather}
where $\A$ is an \emph{assignment map}~\cite{pechukas94a, PhysRevLett.75.3020, embeddingmap, masillo:012101} that captures the mathematical properties of the relationship between the reduced state and the total state. The assignment map captures the essence of the open quantum systems perspective. It represents all the physical assumptions made about the total state as a function of the known reduced system state, containing details about the state of $E$ and $SE$ correlation~\cite{embeddingmap, masillo:012101}. The positivity of $\B$ depends on the interplay of the assignment map $\A$, the details of the unitary evolution, and the averaging of the environment~\footnote{The dynamical map is a stochastic processes that \emph{linearly} maps density matrices into density matrices~\cite{Sudarshan:1961p2145}. It can be written as $\mathcal{B} (\eta_0) =\sum_k \lambda_k C_k \,\eta_0\, C_k^\dagger$, where $\lambda_k$ are the eigenvalues of the map. The trace preservation is imposed by $\sum_k \lambda_k C_k^\dagger C_k=\openone$. Positivity of a map means it maps positive matrices to positive matrices and it is \emph{complete positive} when $\lambda_k \geq 0;\forall\; k$~\cite{choi72a}. The details of the map, $\{\lambda_k\}$ and $\{C_k\}$, depend on the combination of the assignment map, the $SE$ coupling and the trace.}. These three aspects cannot be isolated. Partial trace is a completely positive and linear operation~\footnote{This can be shown by noting that the trace is a map of the form $\mbox{tr}_E \left[ \rho \right] =\sum_e \braket{e |\rho | e} =\eta$ where $\{\ket{e}\}$ forms a complete basis on the space $E$.}, as is the unitary. To completely understand the mathematical properties of the dynamical map, the missing piece is to understand the role and properties of the assignment map.

The assignment map was introduced as a mathematical mapping that takes a matrix in $S$ to a matrix in the $SE$ space~\cite{pechukas94a, PhysRevLett.75.3020}; this is illustrated in Fig.~\ref{arrowdiag}. Refs.~\cite{pechukas94a, jordan:052110} 
show that an assignment map is a linear, positive and consistent~\footnote{Linear assignment maps can be written as Eq.~\eqref{asgmap}. Consistency means this assignment map is the generalized inverse of the trace, as in Fig.~\ref{arrowdiag}, such that $\mbox{tr}_E\left[ \A \left[\eta \right] \right]=\eta$. Positive assignment maps implies that for each density matrix $\eta$, there is a valid total density matrix $\A \left[ \eta \right]=\rho$. This last property was shown to be inconsistent with the no-broadcasting theorem~\cite{embeddingmap} and the Holevo bound~\cite{masillo:012101}.} map if and only if it is of the form $\A_{P}\left[ \eta \right]= \eta \otimes \tau$, where $\tau$ is a density matrix of $E$ (independent of $\eta$)~\cite{Stinespring}, i.e., it has no initial $SE$ correlations. This assignment map is also completely positive, and thus the derived dynamical map is completely positive, independent of the details of the unitary. Conversely, the assignment maps for initially correlated states cannot be linear, positive, and consistent all at the same time. Many researchers have examined how to relaxed the assumption of initial $SE$ product states~\cite{Simmons:1982eu, pechukas94a, buzek, PhysRevLett.75.3020, jordan:052110, Jordan:2006p112} and have proposed physical interpretations for the nonpositivity of the dynamical map. This is important for the practical purpose of doing quantum process tomography for initially correlated $SE$ states, see~\cite{PhysRevA.81.052119, arXiv:1011.6138}. The dynamical role of such correlations and nonpositive maps was shown to be crucial in non-Markovian dynamical maps~\cite{Rodriguez11b, Angelbook, Mazzola12a}. Witnesses for such correlations have been developed \cite{Laine, PhysRevA.84.042113}.

In this article we study the general properties of a dynamical map as a function of the interplay between the system-environment coupling and the assignment map. In the real world, a system has only one particular coupling to the environment. In this paper, we focus on the positivity conditions when an assignment map is combined with a particular unitary evolutions and the trace. We begin with a brief review of assignment maps. From this, we find a formula to determine the positivity of the map that depends on the system-environment coupling and the assignment map. We discuss how this coupling can hide and reveal the nonpositivity of the dynamical map. We prove that it is always possible to construct a specific $U$ that reveals the initial correlations by making $\B$ nonpositive. We also show how the coupling can hide the initial correlations, making the dynamics map positive. Finally, we look at a standard class of Markovian dynamical models, and show how they depend fundamentally on the specific couplings that hide the initial correlations and guarantee positivity.

\emph{Positivity of dynamical maps}.---In~\cite{embeddingmap}, the relationships between $SE$ correlations, linearity, consistency and positivity were summarized using assignment maps defined in terms of a set states, $\{ \mathbb{P}_i\}$, that form a matrix-basis for the space of $S$, i.e., any state of $S$ can be written as a linear (but not convex) sum $\eta_0 = \sum_i r_i \mathbb{P}_i$. Then the assignment is defined as $\A[\eta_0] = \sum_i r_i \mathbb{P}_i \otimes \tau_i$. In this article we will cast the assignment in a different form:
\begin{gather}\label{asgmap}
\A[\eta_0]=\sum_k \alpha_k A_k \,\eta_0\, A^\dag_k
\end{gather}
where $\alpha_k$ are the eigenvalues of the assignment. The condition of consistency is satisfied by demanding  $\sum_k \alpha_k \tr_E[A_k \eta_0 A^\dag_k ] = \eta_0$. The assignment in Eq.~\eqref{asgmap} is equivalent to the assignments given in \cite{embeddingmap, masillo:012101}, see Appendix for a proof.

The assignment takes a density matrix in the $S$ space and maps it to a matrix in the $SE$ space with correlations. For any $\eta_0$ that agrees with the $SE$ correlations then $\A[\eta_0]=\rho_0$. As a technical trick, the state of $E$ is defined to include additional environmental degrees of freedom that are not correlated with the system. Then, the total system-environment state becomes $\rho_0=\Omega_0 \otimes \ket{0}\bra{0}$, where $\ket{0}\bra{0}$ represents the degrees of the environment that are initially uncorrelated with the system, while $\Omega_0$ contains the correlated state. 

\emph{Lemma.} To generate the most general dynamics on $S$ for an arbitrary assignment map, $\A [\eta^S_0] = \rho^{SE}_0$, the total $SE$ state must have the form $\rho^{SE}_0=\Omega^{SE_c}_0 \otimes \ket{0}\bra{0}^{E_r}$. The total space of $E$ is split into two parts: a part that is correlated with $S$ (space $E_c$) and the remaining part that is uncorrelated with $S$ (space $E_r$).

\emph{Proof.} Let the action of the assignment map on $\eta_0$ yield a correlated state of $SE$, $\rho_0$. Now $S$ is not correlated with anything else that it will interact with, if it is then we simply absorb that part into $\rho_0$. The most general dynamics for $S$ then come from the most general dynamics of $\rho_0$, which is a unitary interaction with a pure system, see~\cite{Stinespring, Sudarshan:1986p139} for that proofs. We call the space of $\Omega_0$ to be $SE_c$ and the space of the pure state $E_r$.  Note that $\rho_0$ is not a purification of $\eta_0$. It only contains the systems correlated to $\eta_0$ that will interact with $\eta_0$.$\square$

Combining Eq.~\eqref{asgmap} with Eq.~\eqref{dmap1} gives
\begin{gather}\label{dynmap}
\B(\eta_0)=\sum_{ke} \alpha_k \braket{e| U A_k \; \eta_0 \; A^\dag_k U^\dag| e}.
\end{gather}
The conditions for positivity for the dynamical map is $\braket{s| \B(\ket{r} \bra{r}) |s} \ge 0$ for all $\{\ket{r},\ket{s}\} \in S$. That is if every extremal state of $S$ is mapped to a positive operator, then by convexity every positive operator of $S$ is mapped to a positive operator. The positivity condition in terms of Eq.~\eqref{dynmap} is
\begin{align}\label{finalmap2}
\sum_{ek} \alpha_{k} \braket{se| U A_k |r} \braket{r| A^\dag_k U^\dag |se}
= \sum_k \alpha_k w_{k} \ge 0,
\end{align}
where $w_{k} \equiv \sum_{e} |\braket{se| U A_k |r}|^2$ are positive numbers. The positivity of $\B$ depends on the weighted sum of the eigenvalues of $\A$. Therefore, the values of the weights are important to determine the positivity of $\B$.

The condition for complete positivity is equivalent to finding the eigenvalues of $\B$. From~\cite{Sudarshan:1961p2145} these are found to be
\begin{align}\label{finalmap3}
\sum_{ekrr'ss'} \alpha_k z^*_{rs}z_{r's'}\braket{se| U A_k |r'} \braket{r| A^\dag_k U^\dag |s'e} \ge 0,
\end{align}
where $z_{rs}$ are complex numbers satisfying $\sum_{rs} z^*_{rs} z_{rs} =1$. In general this equation cannot be simplified without specific choices of $\A$ and $U$. Alternatively, we can write Eq.~\eqref{dynmap} as $\B(\eta_0)=\sum_{k} \alpha_k \B_k(\eta_0)$, where $\B_k(\eta_0) \equiv \tr_E [U A_k \eta_0 A^\dag_k U^\dag]$ are non-trace-preserving completely positive super operators. Even though each $\B_k$ is completely positive, the corresponding $\alpha_k$ may not be positive and $\B$ may or may not be completely positive. This is because $\B_k$ are linearly independent, but not simultaneously diagonalizable~\footnote{A non-convex sum of positive operator can also be positive, e.g. $\ket{-}\bra{-} = \ket{0}\bra{0} + \ket{1}\bra{1} - \ket{+}\bra{+}$.}.

What we have shown in Eqs.~\eqref{finalmap2} and~\eqref{finalmap3} is that the positivity and complete positivity of the dynamical map are function of the details of the composition of the assignment map and the unitary dynamics. In the theorem below we give a mathematical construction of interactions $U$ for which $\B$ is nonpositive, provided $\A$ is nonpositive. Then in Eq.~\eqref{cond1} we give a physical condition for the set of interactions $U$ for which $\B$ is always completely positive.

\emph{Theorem.} For every nonpositive assignment there exists some $\eta$ such that $\A[\eta] = \Omega \otimes \ket{0}\bra{0}$, where $\Omega \not\ge 0$. Then there exists a unitary transformation $U$, which leads to nonpositive dynamics for $S$, i.e. there exists $\ket{s}$ such that $\sum_{ek} \alpha_{k} \braket{se| U A_k |r} \braket{r| A^\dag_k U^\dag |se} \not\ge 0$.

\emph{Proof.}---We prove this by explicit construction of a unitary transformation violate the positivity condition given in Eq.~\eqref{finalmap2}, and therefore the condition for complete positivity in Eq.~\eqref{finalmap3} as well.

First note that if the assignment is nonpositive then for a specific state $\eta$ the total state is not positive, and we have $\A [\eta] = \Omega \otimes \ket{0} \bra{0} < 0$. Note $\Omega$ is not positive and therefore not a density matrix. Let us diagonalize this $\Omega \otimes \ket{0} \bra{0}$ in a separable basis~\cite{modigeo}: $\sigma_{1} = U_{1} \Omega  \otimes \ket{0} \bra{0} U_{1}^\dag = \sum r_{ij} \ket{ij}\bra{ij}  \otimes \ket{0} \bra{0},$ where $r_{ij}$ are the eigenvalues of $\Omega$. 

Without loss of generality let us assume that the very first eigenvalue is negative $r_{00}<0$. Although more than one eigenvalues can be negative, we will only need one negative eigenvalue. Next we have $\sigma_{1}= r_{00} \ket{000} \bra{000} + \sum_{j>0} r_{0j} \ket{0j0}\bra{0j0} + \sigma_{\rm rest}$ If we take the trace with respect to $E$, we would get $\eta_{1}= \left(r_{00}+\sum_j r_{0j}\right) \ket{0}\bra{0} + \eta_{\rm rest},$ where $\eta_{\rm rest}=\tr_E[\sigma_{\rm rest}]$. The first eigenvalue of $\eta_{1}$ is $r_{00}+\sum_j r_{0j}$ is a positive number and $\sigma_{\rm rest}$ is a positive operator. Next, apply a control unitary (with $SE_c$ as control) that takes $\ket{0j0}$  to $\ket{0jj}$ for $j>0$ and leaves everything else unchanged. $U_{2}=\ket{00} \bra{00} \otimes \openone +\sum_{j=1}^{d_E-1} \ket{0j} \bra{0j} \otimes v^A_j +\sum_{i=1}^{d_S-1} \sum_{j=0}^{d_E-1} \ket{ij}\bra{ij} \otimes \openone$, where $v^{x}_j=\sum_{k=0}^{d_{x}-1} \ket{k+j} \bra{k}$. The state after this transformation is $\sigma_{2}= U_{2} \sigma_{1} U_{2}^\dag =r_{00} \ket{000} \bra{000} + r_{0j} \ket{0jj} \bra{0jj} + \sigma_{rest}$. After this, apply a control unitary with $E$ as control $U_{3}= \openone \otimes \ket{00} \bra{00} + v^S_j\otimes \sum_{j=0}^{d_E-1} \otimes \ket{jj} \bra{jj} + \openone \otimes \sum_{j \neq k} \ket{jk}\bra{jk}.$ The state after this transformation gives the desired result. $\sigma_{3} = U_{3} \sigma_{2} U_{3}^\dag= r_{00} \ket{000} \bra{000} +r_{0j} \ket{jjj} \bra{jjj} + \sigma_{rest}$ Taking the partial trace with respect to $E$ we get $\eta_{3}=r_{00} \ket{0}\bra{0}+\sum_j r_{0j} \ket{j}\bra{j}+\eta_{\rm rest}$.
All $r_{0j} \geq 0$ and $\eta_{\rm rest}$ is a positive operator that does not contain the matrix $\ket{0}\bra{0}$. And because $r_{00}<0$ we have $\eta_{3}<0$. 

We now consider the following dynamical map from Eq.~\eqref{dynmap}. We let $\A[\eta]=\Omega \otimes \ket{0}\bra{0}$ and $U=U_{3}U_{2}U_{1}$. This map will violate the positivity condition in Eq.~\eqref{finalmap2} in the main text when $\ket{s}=\ket{0}$. This proves that for a nonpositive assignment there exists a dynamical process that leads to not completely positive dynamical map. $\square$

Pechukas~\cite{pechukas94a} showed that if there are any initial correlations in $SE$ then the assignment map is nonpositive. Here we have shown that the nonpositivity of this assignment map can always be revealed as nonpositive of the dynamics of $S$ given an appropriate unitary transformation. The unitary we constructed in the proof is one such transformation, there can be many others.

Now that we have shown how to reveal nonpositivity of $\A$ in the dynamics of $S$, we show how it can be hidden. For that we exploit the bipartite decomposition: $\rho = \eta \otimes \tau + \chi$, where $\chi$ is the correlations matrix~\cite{Carteret08a}. Note that any bipartite state can be written in this form and $\tr_S[\chi]=\tr_E[\chi]=0$. The correlation matrix has physical importance as it links the states of $S$ and $E$. Our physical condition and subsequent interpretation rely on this matrix.

We remark that the set of unitary transformations $\{ W \}$ satisfying
\begin{gather}\label{cond1}
\tr_E\left[W \chi_0 W^\dag \right]=0
\end{gather}
lead to completely positive dynamics. This can be seen by noting that the action of the dynamical map is $\B(\eta_0)= \tr_E \left[W \left\{ \eta_0 \otimes \tau_0 + \chi_0 \right\} W^\dag \right] = \tr_E \left[ W \eta_0 \otimes \tau_0 \; W^\dag \right] +\tr_E \left[W \chi_0 W^\dag \right]$. When the second terms is vanishing we have $\B(\eta_0)=\tr_E\left[W \eta_0 \otimes \tau_0 \; W^\dag\right]$, which is completely positive~\cite{Stinespring, Sudarshan:1986p139}.

The authors of~\cite{Hayashi} investigated the unitary transformations that always lead to completely positive dynamics for any correlations; the answer turns out to be the local unitary transformation, $U=U_S \otimes U_E$. This can be seen as a direct consequence of the Eq.~\eqref{cond1} above since $\tr_E[(U_S \otimes U_E) \chi_0 (U_S \otimes U_E)^\dag] = \tr_E[U_S \chi_0 U_S^\dag] = 0$.  We will now see the implications of Eq.~\eqref{cond1} as it applies to models of Markovian dynamics.


\emph{Markovian Models.}---In order to highlight the significance of Eq.~\eqref{cond1}, we will focus on its role within decoherence models that rely on environmental refreshing~\cite{rau, PhysRevLett.88.097905, Bruneau, PhysRevA.77.032121}. A refreshing model is one where $S$ periodically interacts with a part of $E$, $\tau_n$, for duration time $T$. The total state of $E$ is $\tau = \tau_0 \otimes \tau_1 \otimes \tau_2 \dots \otimes \tau_n \otimes \dots.$ The $SE$ interactions come from a unitary of the form $U_t=\exp\left[-i t H_t \right]$ where the time dependent Hamiltonian is $H_t=\sum_n \theta(t,T,n)\;V_{n}$ where 
\begin{align}
\theta(t,T,n)=\left\{   \begin{array}{cc}
\mbox{if } nT\leq t\leq (n+1)T:& 1        \\
\mbox{for all other } t:   &  0   
\end{array} \right.
\end{align}
and $V_{n}$ is a Hamiltonian that couples $\eta$ to $\tau_n$. Furthermore, it is often assumed that each interaction $V_{n}$ is identical to each other, except that they act on a different state $\tau_n$. Such an unitary couples $S$ in an identical fashion to different parts of $E$ every $t=nT$. Thus, the evolution of a step of $S$ is given by: $\eta_{n+1} = \tr_E \left[ e^{-iT V_{n}} \eta_{n} \otimes\tau_n e^{iT V_{n}} \right] =\B \left( \eta_{n} \right).$ The repeated action of such a map can be written as $\eta_{n+1}=\B^n\left( \eta_{0} \right).$ This is a quantum version of the Boltzmann collision model of the ideal gas. These models have been shown to have thermalization properties similar to the Markovian master equation for timescales much larger than $T$~\cite{PhysRevLett.88.097905, Bruneau}.

To understand how such a model deals with the $SE$ correlations $\chi$, we will now examine the behavior of $\chi$ for one refreshing step. At $t=0$, $\rho_0 = \eta_0 \otimes \tau_0$. Thus, $\chi_{0}=0$. After coupling $S$ and $E$ for some time $t=T$, correlations between $\eta_1$ and the part of $E$ will have developed, giving rise to a $\chi_1 \neq 0$. However, due to the nature of the coupling of the refreshing model, such correlations will not have an impact on later steps. Note that for the next step, $\eta_1$ will be coupled to $\tau_1$, making $\tr_E \left[U_1 \chi_1 U_1^\dag\right]=0$. Similarly, for each step, the correlations are discarded $\tr_E \left[ U_{n} \chi_{n} U_{n}^\dag\right]=0$. Eq.~\eqref{cond1} shows how these Markovian models are completely positive.

\emph{Conclusion}.---We have found the conditions for positivity for dynamical maps coming from correlated system-environment ($SE$) states. These correlations can sometimes make the dynamical maps nonpositive, which make their use difficult. Thus, finding if a map is positive simplifies its use. We used linear assignment maps that can create $SE$ correlations, and considered the most general $SE$ couplings. Similarly, we have found the conditions for complete positivity of the map.

We showed how the positivity of the map depends on the interplay between the assignment map and the $SE$ coupling. For correlated states the assignment map can be non-positive, and still have a meaningful physical interpretation. The specific of the $SE$ coupling can hide or reveal this non-positivity, affecting in turn the positivity of the dynamical map. We prove that if the assignment map has negative eigenvalues, there always exists a $SE$ coupling that will reveal this negativity by making the dynamical map non-positive. We show how to construct such a coupling.

The $SE$ coupling can also hide the negativity of the assignment map. We give an expression for the conditions that the $SE$ coupling, when fulfilled, the $SE$ correlations are hidden making the dynamical map completely positive. We show how a very large class of Markovian models, known as refreshing models and Boltzmann collision models, are completely positive and Markovian precisely because their couplings are chosen to periodically hide the $SE$ correlations. 

These results highlight the dynamical role of positive and non-positive maps in physically-motivated open quantum systems. This formulation explains how to use assignment maps to expand the dynamical map formalism to account for initial correlations and non-Markovian effects, expanding its utility. At the same time, these results explains the role of system-environment correlations in many commonly used models.

{\bf Acknowledgements.} KM is supported by the John Templeton Foundation, the National Research Foundation, and the Ministry of Education of Singapore. KM thanks the Department of Chemistry and Chemical Biology at Harvard University for hospitality. CAR thanks the Centre for Quantum Technologies for their hospitality.

\appendix

\section*{Appendix}

The assignment presented in~\cite{embeddingmap} is of the form $\A[\mathbb{P}_i]= \mathbb{P}_i \otimes \tau_i=\mathbb{R}_i$, where $\{\mathbb{P}_i\}$ form a linearly-independent matrix basis on the space of $S$, i.e. any state of $S$ can be written as $\eta=\sum_i r_i \mathbb{P}_i$.
\begin{gather}\label{appass}
\A \left[ \sum_i r_i\mathbb{P}_i \right]= \sum_i r_i \A[\mathbb{P}_i]= \sum_i r_i \mathbb{R}_i.
\end{gather}
The consistency condition requires $\tr_E[\mathbb{R}_i]=\mathbb{P}_i$ (and therefore $\tr[\mathbb{R}_i]=1$). Additionally, Hermiticity preservation requires that $\mathbb{R}_i=\mathbb{R}_i^\dag$. Note above, $\{\mathbb{P}_i\}$ are density operators but $\{\mathbb{R}_i\}$ are not necessarily positive. Here we show that this is the same as a map in Eq.~\eqref{asgmap} in the main text.

\emph{Lemma 2.} For any set of Linearly independent matrices $\{\mathbb{P}_i\}$, there exists the dual set $\{\Delta_i\}$ satisfying $\tr[\Delta_i \; \mathbb{P}_i] =\delta_{ij}$.

\emph{Proof.} Write $\mathbb{P}_i = \sum_j h_{ij} \Gamma_j$, where $h_{ij}$ are real numbers and $\{\Gamma_j\}$ form a Hermitian self-dual linearly independent basis satisfying $\tr[\Gamma_i \Gamma_j]=2 \delta_{ij}$~\cite{PhysRevA.68.062322}. Since $\{\mathbb{P}_i\}$ form a linearly independent basis, the columns of matrix $\mathsf{H} = \sum_{ij} h_{ij} \ket{i}\bra{j}$ are linearly independent vectors, which mean $\mathsf{H}$ has an inverse. Let matrix $\mathsf{D}^T=\mathsf{H}^{-1}$, then $\mathsf{H} \mathsf{D}^T = \mathsf{I}$, implying that the columns of $\mathsf{D}$ are orthonormal to the columns of $\mathsf{H}$. We define $\Delta_i = \frac{1}{2} \sum_j d_{ij} \Gamma_j$, where $d_{ij}$ are elements of $\mathsf{D}$. $\square$

\emph{Lemma 3.} A map in the form of Eq.~\eqref{appass} is equivalent to the map of the form Eq.~\eqref{asgmap} in the main text.


\emph{Proof.} We write the map in Eq.~\eqref{appass} as
\begin{gather}\label{appassmap}
\A[\eta]=\sum_{i} \tr[\Delta_i \; \eta] \; \mathbb{R}_i.
\end{gather}
First note that by this construction Eq.~\eqref{appassmap} satisfies Eq.~\eqref{appass}. Next, we can write the operators $\mathbb{R}_i$ and $\Delta_i$ in their eigenbasis:
\begin{align}
\A[\eta]=&\sum_{im} \tr \left[d_{im} \ket{d_{im}} \bra{d_{im}} \eta \right] \sum_{in}\ket{r_{in}}\bra{r_{in}}\\
&\sum_{imn} d_{im} r_{in} \ket{r_{im}} \bra{d_{in}} \eta \ket{d_{im}} \bra{r_{in}}
\end{align}
Next we define $\alpha_k=d_{im}r_{in}$ and $A_k= \ket{r_{im}} \bra{d_{in}}$ and we have the desired from. 

Conversely, to cast the map in the form of Eq.~\eqref{appass}, we have to chose a set of linearly independent matrices as the basis. The action of the map in Eq.~\eqref{asgmap} in the main text acting on the elements of the linearly independent basis gives us $\mathbb{R}_i=\sum_k \alpha_k A_k \mathbb{P}_i A^\dag_k$. $\square$

Through out this Letter, we are use a different notation for assignment maps than in~\cite{embeddingmap}. To aid the reader, we will prove that the assignment maps from~\cite{embeddingmap} can always be written as in Eq.~\eqref{asgmap} in the main text. The proof is as follows. In ~\cite{embeddingmap}, the assignment map was written as \begin{gather}\label{oldpaperassignment}
\A[\eta]=\sum_{j} \tr[\Delta_j \, \eta] \; \mathbb{P}_j\otimes\tau_j,
\end{gather}
which is clearly of the form of Eq.~\eqref{appassmap}. Note that $\eta$, $\mathbb{P}_j$ and $\Delta_j$ are matrices in the space of $S$ and, while $\tau_j$ are matrices in the space of $E$. Note that $\tr[\Delta_j \eta] \mathbb{P}_j$ can be expanded using an additional index $m$ such that $\tr[\Delta_j \eta] \mathbb{P}_j=\sum_m \mu_{m,j} M_{m,j} \,\eta \,M_{m,j}^\dagger $. Also, $\tau_j$ can be expanded on its eigenbasis $\{ \ket{T_{n,j}} \}$ such that $\tau_j=\sum_n t_{n,j} \ket{T_{n,j}} \bra{T_{n,j}}$, where $n$ runs up to $e$. Thus, 
\begin{gather}
\A[\eta]=\sum_{j}\sum_{m,n} \mu_{m,j} t_{n,j} M_{m,j}\, \eta\, M_{m,j}^\dagger\otimes \ket{T_{n,j}} \bra{T_{n,j}} \nonumber
\end{gather}
This can be cast on the form of Eq.~\eqref{asgmap} in the main text by combining the indices $k=\{j,m,n\}$ such that $\alpha_k=\mu_{m,j}t_{n,j}$ and $A_k= M_{m,j}\otimes\vert T_{n,j}\rangle$. Note that $A_k$ is a rectangular matrix, mapping from $S$ space to the $SE$ space. This proves how to write Eq.~\eqref{oldpaperassignment} in the form of Eq.~\eqref{asgmap} in the main text.

\bibliography{CPUnitary-3}

\end{document}